\documentclass[article]{JHEP}
\usepackage{amsmath,epsfig}
\usepackage{amssymb,amsfonts}
\title{Five-Brane configurations without\\
a strong coupling regime} \preprint{hep-th/0204201\\LPTENS/02-28\\
CPTH-S036.0801}
\author{E. Kiritsis$^{\ 1,2}$, C. Kounnas$^{\ 1}$,
P.M. Petropoulos$^{\ 3}$ and J. Rizos$^{\ 4}$\\
$^{\bf (1)}$ Laboratoire de Physique Th{\'e}orique
de l'Ecole Normale Sup{\'e}rieure\footnote{Unit{\'e} mixte  du
CNRS et de l'Ecole Normale Sup{\'e}rieure, UMR 8549.}\\
24 rue Lhomond, 75231 Paris Cedex 05, FRANCE\\
$~$\\
$^{\bf (2)}$ Department of Physics, University of Crete, and FO.R.T.H.\\
71003 Heraklion, GREECE\\
$~$\\
$^{\bf (3)} $ Centre de Physique Th{\'e}orique,
Ecole Polytechnique\footnote{Unit{\'e} mixte  du CNRS et de  l'Ecole
Polytechnique,
UMR 7644.}\\
 91128 Palaiseau Cedex, FRANCE\\
$~$\\
$^{\bf (4)} $ Physics Department, University of Ioannina\\
45110 Ioannina, GREECE\\}

\abstract{Five-brane distributions with no strong-coupling
problems and high symmetry are studied. The simplest configuration
corresponds to a spherical shell of branes with $S^3$ geometry and
symmetry. The equations of motion with $\delta$-function sources
are carefully solved in such backgrounds. Various other brane
distributions with sixteen unbroken supercharges are described.
They are associated to exact world-sheet superconformal field
theories with domain-walls in space--time. We study the equations
of gravitational fluctuations, find normalizable modes of bulk
six-dimensional
gravitons and confirm the existence of a mass gap. We also study
the moduli of the configurations and derive their (normalizable)
wave functions. We use our results and holography to calculate,
in a controllable fashion, the two-point function of the stress
tensor of little string theory in these vacua.}


\newcommand{\p}{\partial}

\newcommand{\nn}{\nonumber}

\def\bea{\begin{eqnarray}}
\def\eea{\end{eqnarray}}
\newcommand{\be}{\begin{equation}}
\newcommand{\ee}{\end{equation}}

\def\a{\alpha}
\newcommand{\ba}{\begin{eqnarray}}
\newcommand{\ea}{\end{eqnarray}}
\def\p{\partial}
\def\e{\, {\rm  e }}
\def\mysquare
{\large\hbox{{$\sqcup$}\llap{$\sqcap$}}}
\def\R{{\bf R}}

\def\np#1#2#3{Nucl. Phys. {\bf{B#1}} (#2) #3}
\def\pl#1#2#3{Phys. Lett. {\bf{B#1}} (#2) #3}

\begin{document}
\maketitle
\section{Introduction and summary}

The notion of branes and a successful description of their dynamics
has proven to be very fruitful both for understanding the fundamentals of
string/M-theory, and in order to investigate non-trivial vacua of the theory
that may describe observable low-energy physics. At the fundamental level,
BPS branes (NS5-branes \cite{kounnas,CHS}, D-branes \cite{polchinski}) are
essential for the unification of various perturbative vacua of string
theory under the umbrella of M-theory \cite{witten}. In addition, they have
provided profound connections between gauge theory (dynamics of
fluctuations) and gravity (dynamics of long-range bulk fields), leading to
brane-engineering of field theories \cite{hw} and precise formulations of
bulk-boundary (holographic) correspondence \cite{maldacena}. Moreover, they
have provided, especially via orientifolds, many new examples of vacua that
seem very promising for describing real physics with a string scale that
may  be accessible to experiment
\cite{wla}.

An especially interesting and also untamed type of brane is the magnetic
dual of the fundamental string, namely the NS5-brane. It is a BPS object
breaking half the supersymmetry of the original theory. All closed string
theories contain an NS5-brane.  The world-volume theory depends on the
type of the parent string theory. The type IIB NS5-brane world-volume
theory has $(1,1)$ supersymmetry in six dimensions (16 supercharges) and is
non-chiral. Its massless spectrum is a vector multiplet. It contains in
particular four scalars that are the goldstone modes of the (broken)
translational invariance in the four transverse dimensions. The full
world-volume theory is a string theory, known as little string theory (LST).
By utilizing the S-duality of the theory, the NS5-brane is mapped to the
D5-brane which has the conventional Polchinski description in terms of open
strings. The little strings can be thought of as the intersection
points of a D3-brane ending
on a IIB NS5-brane. When we have $N$ coinciding
five-branes we expect symmetry enhancement and zero-mass charged gauge bosons
on the branes. NS5-brane vacua have also been conjectured to describe the
high-temperature behavior of string theory \cite{adk}.

The NS5-brane of type IIA theory, has a chiral world-volume theory with
$(2,0)$ supersymmetry. It is the direct descendant of the M-theory
five-brane, which is describing the strong-coupling limit of the type IIA
NS5-brane. It has its own world-volume LST. The massless
spectrum is a tensor hyper-multiplet, containing a self-dual two-index
antisymmetric tensor, and five scalars with a similar interpretation as in
type-IIB case. In this case, the object that can end on the NS5-brane is a
D2-brane. Its intersection is a
string that is minimally charged under the world-volume self-dual tensor.
There is a decoupling limit $g_{\rm s}\to 0$, where the interactions of the
world-volume fluctuations decouple from the bulk \cite{Seiberg:1997zk}.
In this limit,
the world-volume theory is a non-critical string theory with length scale
$\ell_{\rm s}=\sqrt{\alpha'}$ and no dimensionless coupling. This is a strongly
coupled
string theory about which we know very little. It is the mother of the only
non-trivial fixed-point field theories known in six dimensions. At distances
much larger than $L\equiv \sqrt{\alpha' N}$ the theory is effectively a
non-trivial $(2,0)$ superconformal field theory.

Symmetry enhancement is also expected here, when we have $N$ branes
coinciding in transverse space. This is however more exotic that type IIB
since here it is the D2-branes stretching between the NS5-branes that become
tensionless in the coincidence limit, implying that their boundary strings
are tensionless. This is a generalization of the Higgs mechanism of gauge
theories to a theory of self-dual antisymmetric tensors. We do not have
yet a good understanding of this effect.

These solitons correspond to supergravity solutions with non-trivial metric,
dilaton and antisymmetric tensor \cite{CHS}. In their near-horizon region
the solution has an exact conformal field theory (CFT) description
\cite{kounnas,K93,AFK}
in terms of $SU(2)_k\times{U(1)_Q}$ plus free fermions  with a linear dilaton.
Such a solution describes a collection of $N=k+2$ NS5-branes
located at the same
point \cite{CHS}. These solutions have the property that the effective
string coupling, $\exp{\phi},$ diverges at the location of the five-brane. This
renders problematic the string description of effects associated with the modes
localized on the brane.

The near-horizon limit is the decoupling limit described above. Thus, it is
expected that holography might be at work also here \cite{Itzhaki:1998dd,
ABKS}. The claim is that supergravity in the $SU(2)_k\times{U(1)_Q}$ in the
limit of large $N$, is holographically dual to the LST. As in the
usual AdS/CFT correspondence, one expects to learn
more from such a duality both
for the gravity side as well as for the LST side. To apply however the
techniques of holography one needs a controlled supergravity/string theory
description of the bulk theory, and this is seriously hampered by the fact
that the effective coupling, parameterized by the background dilaton, is
strong in some regions
of space--time. This strong-coupling problem is not new in string theory,
with a prototype being Liouville theory.
The way around has been to somehow
modify the theory so that the strong-coupling region is ``screened".
This can be
achieved either by cutting it off by fiat\footnote{This would correspond
in the case of the $c=1$ string theory to passing from a Liouville theory to
the $SL(2,{\bf R})/U(1)$ coset CFT.}, or by modifying the theory so that it is
dynamically disfavored for the system to go near the strong-coupling
region. In the case again of
Liouville this amounts to adding a potential that screens off the
strong-coupling region. Experience from $c=1$ string theory suggests that the
un-regularized linear dilaton background is singular.

In the case of the supergravity description of five-branes we are faced
with a similar problem. Several attempts have been made to regularize the
strong-coupling behavior. One approach, \cite{ABKS} (anticipated in
\cite{Itzhaki:1998dd}) is to replace the standard type IIA NS5-brane, in the
strong-coupling region (near the brane) by its eleven-dimensional ancestor,
the M5-brane. This can be achieved by starting from a solution of M-theory
describing M5-branes distributed on the M-theory circle. At short distances
the M-theory circle is large, but it asymptotically goes to zero, producing the
NS5-brane solution of the type IIA theory. This is an elegant approach, but
its down-turn is that the metric is complicated especially in the intermediate
region, and a successful application of holography requires mastering the
geometrical data well.

A different attempt has been to consider NS5-branes distributed uniformly on a
circle in transverse space \cite{Giveon:1999zm}. In \cite{Sfetsos:1999xd} it
was observed that such a distribution, in the continuum limit, is T-dual to
the geometry  of a $Z_k$ orbifold of the
$SL(2,{\bf R})_{k+4}/U(1)\times{SU(2)_k/U(1)}$
coset CFT. This dual coset space regulates
the strong coupling \cite{K93}. With this
starting point, several holographic issues of such a
distribution have been analyzed \cite{Giveon:1999zm,Gava:2002gv}. The
picture in terms of five-branes on a circle may be an
oversimplification. In a curved non-compact background, T-duality may
\cite{Kiritsis:1993ju,Giveon:1994ph} or may not be an exact symmetry.
An NS5-brane with one longitudinal direction wrapped on a
circle is T-dual to flat space \cite{kkl}, although we have serious reasons
to believe that the dynamics in this case is non-trivial. The Nappi--Witten
pp-wave background \cite{Nappi:1993ie}, which is also T-dual to flat space
\cite{kk}, is not equivalent to flat space or a standard orbifold of
it, and this can be asserted
since its exact solution is known \cite{kk}.

In this work we will investigate other NS5-brane distributions, that have
the
property that the strong-coupling region is absent, and they have high
symmetry
so that detailed calculations become possible. Continuous distributions of
branes and in particular five-branes have been studied before
\cite{Sfetsos:1999xd,BS,Bakas:2000ax,Kraus:1999hv,kounn,Giddings:2000zu,
Martelli:2002tu}.
A characteristic of the distributions we will use (which are infinitely thin
shells) is that they generate a discontinuous geometry and they need the
inclusion of sources. However, as we explicitly show, they are controllable
backgrounds, and the study of small fluctuations around such backgrounds is
well defined.

One of our aims is to consider distributions that correspond to exact
conformal
field theories albeit of a new kind. They correspond to sowing together (in
space--time instead of the world-sheet \cite{Bachas:2001vj}) known CFT's. The
simplest example is a spherical shell of $N$ NS5-branes distributed uniformly
on
an $S^3$ in transverse space. The number $N$ should be large enough
so that the geometry is weakly curved, and therefore $\a'$ corrections to
supergravity negligible.
Large $N$ also ensures that the brane distribution can be approximated
by a continuous one and consequently enjoy high symmetry ($SO(4)$).

In the interior of the shell the geometry and other background fields are
flat.
In that sense, this is somewhat reminiscent of the enhancon configuration
\cite{Johnson:2000qt}. There are five-brane $\delta$-function sources at the
position of the shell, which are determined uniquely from the supergravity
equations, as we show. The radius of the shell, $R$, can be chosen large
$R\gg\sqrt{\alpha' N}$ so that the string coupling is weak outside the
shell. Inside the
shell the string coupling is frozen. Hence, there is no strong-coupling
region in such a background.

A richer variety of such backgrounds can be achieved by also using
negative-tension
branes. In the case of the D5-branes these are no other than the orientifold
five-planes. For NS5-branes, their negative-tension cousins are ``bare" $Z_2$
orbifold five-planes. A usual orbifold five-plane appearing as a
twisted sector in
closed-string orbifold vacua is a bound state of an NS5-brane and a bare
orbifold plane that cancels the tension and charge of the NS5-brane much
alike the situation in orientifold vacua. The twisted-sector fields are the
fluctuations of the NS5-brane since negative-tension branes have
no fluctuations in string theory (because of unitarity).

In such backgrounds one can study the spectrum of fluctuating fields. These
should correspond via holography to operators of the boundary LST. The
effective field theory of such fluctuations is expected to be a
seven-dimensional, $SO(4)$ gauged supergravity. It is  obtained by
compactifying the ten-dimensional type-IIA/B supergravity (in the string
frame) on $S^3$ with the appropriate parallelizing flux of the antisymmetric
tensor.  The vacuum corresponding to the near-horizon region of an
NS5-brane should
correspond to a flat seven-dimensional space plus a linear dilaton in one
direction. This is expected to be the holographic direction. To our knowledge,
this gauged supergravity in seven dimensions has not
yet been constructed. However, other $SO(4)$ gauged supergravities are  known in seven and
four dimensions \cite{AFK}.

In the present paper,  we will solve explicitly for the
fluctuations of some of the
fields of the bulk theory. These include the six-dimensional graviton
(corresponding to the boundary stress tensor) and its associated
Kaluza--Klein (K--K) tower.
It turns out that the six-dimensional graviton satisfies an equation without
sources. We find the normalizable modes and show that its spectrum has a
mass gap $\sim {{\ell+1}\over \sqrt{N}}M_{\rm s}$. This was expected
from an earlier CFT
computation \cite{KK}. The modes under consideration correspond to
long representations of $N=2$
supersymmetry in six dimensions \cite{kounn}.

The other set of fluctuations we consider are the moduli modes which are
massless (short representations of $N=2$). These satisfy a Laplace equation
with sources \cite{kounn}. The sources are
crucial for the existence of normalizable moduli modes, as we show.

We further study the non-normalizable modes of the six-dimensional graviton
in order to apply the holographic principle. The symmetries of the background
we are studying are $SO(4){\times}ISO(6)$. The $SO(4)$ corresponds to
the  R-symmetry
of the boundary theory, while the rest is the usual Euclidean group in six flat
dimensions. This is unlike AdS-like spaces where conformal
transformations are also boundary symmetries.

Using the bulk supergravity action, we can compute the boundary two-point
function of two stress tensors. It has the following features:

\noindent
(\romannumeral1) its long distance behavior is massive with associated mass
${M_{\rm s}\over \sqrt{N}}$;

\noindent
(\romannumeral2) in the formal $N\to\infty$ limit it becomes power-like with a
$|x|^{-7}$ behavior;

\noindent
(\romannumeral3) the stress tensor has canonical mass dimension 7 due to a
non-trivial IR wave-function renormalization of its source;

\noindent
(\romannumeral4) it is independent of
the presence of the shell and, as we argue, this is no longer true for
higher correlators.

The structure of this paper is as follows. In Section 2 we review the
standard
five-brane solutions. In Section 3 we find the five-brane
distributions that
we use as solutions of the supergravity equations with sources. In Section 4 we
describe similar solutions for orientifold and orbifold five-planes.
Section 5 contains an analysis of elaborate
distributions of five-branes and five-planes with $SO(4)$ symmetry,
all of the same kind, i.e. either all charged under NS--NS or all under R--R.
In Section~6 we describe a solution that interpolates between D5 and
NS5-branes in type IIB string theory. In Section 7
we study the fluctuation spectrum of the six-dimensional graviton.
In Section 8
we discuss holography and calculate the two-point function of the
stress tensor.
In Section 9 we investigate the moduli of the configuration and calculate
their
normalizable wave functions. Finally, Section 10 contains our conclusions and
further problems. In the appendix we present the bulk-to-bulk propagator in
the background under investigation.

\boldmath
\section{The dilatonic five-brane solutions: a reminder}\label{5b}
\unboldmath

The canonical five-brane solutions have been extensively studied
in the literature. They are determined by minimizing the
ten-dimensional effective action which reads, in the Einstein frame:
\begin{equation}
S^{(10)}=\frac{1}{2\kappa^2_{10}}\int d^{10}x \, \sqrt{-g^{(10)}}\left(
R^{(10)}-\frac{1}{2}(\partial\phi_\gamma)^2-\frac{1}{12}
\e^{-\gamma\phi_\gamma}\,H_{\gamma}^2\right).
\label{act}
\end{equation}
Here $\phi_\gamma$ is the dilaton field and $\gamma=\pm 1$ corresponds to
the
two distinct NS--NS or R--R three-form field strengths $H$ in type IIB
theory (type IIA allows only for $\gamma=+1$). We do not introduce
any gauge field,
which means in particular that the branes under consideration carry no other
charge than NS--NS or R--R. Notice that the ten-dimensional Newton's
constant
appears in $2\kappa^2_{10}\equiv 16\pi G^{\vphantom 2}_{10}= (2\pi)^7
{\alpha'}^4$.

We seek for solutions of the type
\begin{equation}
\frac{ds^2}{\alpha'}=a(z)\left(-dt^2+d\vec{x}^2\right)+ b(z) \e^{2 z}
\left(dz^2 + d\Omega^2_3\right),
\label{5met}
\end{equation}
where $\vec{x}$ are Cartesian coordinates in a five-dimensional Euclidean
flat
space and
\begin{equation}
d\Omega^2_3=d\theta^2+\sin^2 \!\theta\left(d\varphi^2+ \sin^2
\!\varphi\,d\omega^2\right)
\label{3sph}
\end{equation}
is the metric on a unit-radius three-sphere. Together with $z$, the latter
is
transverse to the five-brane, and $r=\exp z$ is the radial (dimensionless)
transverse coordinate. Poincar{\'e} invariance within the
five-brane world-volume is here
automatically implemented. For canonical five-brane solutions, we must also
assume that the functions $a(z)$, $b(z)$, as well as the dilaton
$\phi_\gamma
(z)$ are expressed in terms of a single {\it positive} function, $h(z)$: \ba
a(z)&=&{h(z)}^{-1/4},\label{aeq}\\
b(z)&=&{h(z)}^{3/4},\label{beq}\\
\phi_{\gamma}(z)&=&\frac{\gamma}{2}\log h(z).\label{deq} \ea
 Moreover, the three-index antisymmetric tensor lives on $S^3$:
\begin{equation}
\frac{H}{\alpha'} = 2 f(z) \sin^2 \!\theta \, \sin \varphi\, d\theta
\wedge d\varphi \wedge d\omega.
\label{3form}
\end{equation}
The function $f(z)$ must be piece-wise constant in order to ensure $dH=0$
except at the location of the branes which act like sources.

With the above ansatz (Eqs. (\ref{5met})-(\ref{3form})), we can readily
solve
the equations of motion of (\ref{act}). We find:
\begin{equation}
f=
-\frac{h'}{2} \e^{2z}
\label{feq}
\end{equation}
with $h(z)$ a harmonic function satisfying
\begin{equation}
\mysquare_{}  h = 0.
\nn
\end{equation}
The general solution is therefore \ba h(z) = h_0+N \e^{-2z}, \label{heq}
\ea
with $N$ and $h_0$ two integration constants, which are both positive for
$h(z)$ be positive. The first one, $N\geq 0$, is interpreted as the total
number of five-branes, sitting at $z \to -\infty$ ($r=0$). According to Eq.
(\ref{feq}),
\begin{equation}
f(z)=N,
\nn
\end{equation}
in this case. If no five-branes are present, we recover flat-space with
constant dilaton and no antisymmetric tensor. If $h_0=0$, the transverse
geometry is an $S^3$ of radius $L=\sqrt{\alpha' N}$ with a covariantly
constant antisymmetric tensor (proportional to the three-sphere volume form)
plus a linear dilaton (we have introduced $y = z \sqrt{N}$, and
$\psi_\gamma
(y) \equiv \phi_\gamma (z)$):
\ba
\frac{ds^2}{\alpha'}&=&\e^{-\frac{\gamma}{2}
\psi_\gamma (y)
} \left(-dt^2+d\vec{x}^2 +dy^2 + N\, d\Omega^2_3\right),
\label{nhg}\\
\psi_{\gamma}(y)&=&\frac{\gamma}{2}\log N
-\frac{\gamma y}{\sqrt{N}}. \label{lidi} \ea

The solution at hand, Eq. (\ref{heq}), is the neutral five-brane of
\cite{CHS}.
The space is asymptotically flat: when $z\to +\infty$, i.e. $r\to \infty$,
the
dominant term in (\ref{heq}) is the constant $h_0$. On the other hand, the
limit $z\to -\infty$ corresponds to the {\it near-horizon geometry}, $r\to
0$,
where $h_0$ is negligible and the geometry approaches (\ref{nhg}) with
linear
dilaton (\ref{lidi}). As far as the string coupling is concerned, from Eqs.
(\ref{deq}) and  (\ref{heq}), we learn the following: (\romannumeral1) when
$\gamma = + 1$ (NS), $g_{\rm s}$ diverges at $r \to 0 $ and is bounded from
below
by $h_0$ at $r \to \infty$; (\romannumeral2) when $\gamma = - 1$ (D),
$g_{\rm
s}$ vanishes at $r \to 0 $ and is bounded from above by $1/h_0$ at $r \to
\infty$, except for the special case $h_0 = 0$.

The situation with $h_0 = 0$, described in (\ref{nhg}) and  (\ref{lidi}), is
of
particular interest. Considered as a bulk type II geometry, the latter is an
 exact $N=4$ superconformal theory \cite{kounnas,CHS,K93,AFK}.
In the case $\gamma = + 1 $ (NS), this theory is a two-dimensional
 $\sigma$-model,
whose target space is the ten-dimensional manifold,
$K^{10}_{\vphantom k}\equiv W_{k}^{4}\times{M^6_{\vphantom k}}$.
Here $M^6$ is a flat six-dimensional space--time and
$W_{k}^{4} \equiv U(1){\times}SU(2)_{k}$, the four-dimensional background
with linear dilaton. The $N=4$ superconformal symmetry implies, for type II
strings, the existence of $N=2$ space-time supersymmetry in six
dimensions (1/2 of
the initial supersymmetry). In this background, as we have already pointed out,
the string coupling constant becomes infinitely large at the location of the
NS5-brane and the {\it string perturbation brakes down.} Notice that for the
D5-brane background ($\gamma = -1$), the same phenomenon occurs at $z \to
+\infty$, i.e. in the asymptotic region, far away from the sources.

Many proposals can be found in the literature, which aim to properly define
the
above string theory in that region of space--time where its coupling
diverges.
In the type IIB string, one can advocate S-duality which turns large
coupling
into small  $g_{\rm s}\to 1/g_{\rm s}$ and the NS5- to D5-brane. At the
location of the D5-brane, the bulk coupling constant goes to zero. In this
representation, one can use type-IIB- and open-string-theory techniques to
study the D5-brane dynamics decoupled from the bulk. Similarly, for type
IIA,
duality lifts us to eleven-dimensional M-theory, where one deals with
M5-branes.

Another way to handle the above large-coupling pathology is based on
T-duality, by replacing the $W_{k}^{4}$ background with a T-dual,
four-dimensional space, and identical $N=4$ superconformal symmetries
\cite{adk,K93,AFK}:
\begin{equation}
\Delta_{k}^{4}= \left(\frac{SU(2)}{ U(1)}\right)_k \times
\left(\frac{SL(2,{\bf
R})}{ U(1)_{\rm axial}}\right)_{k+4}.
\nn
\end{equation}
In this expression, both factors are exact CFT's based on gauged
WZW models. The first is described by compact parafermions, while
the second is the two-dimensional Euclidean black hole constructed
as the axial gauging of $SL(2,\R)$.  The important fact here is
that the value of the string coupling (in the axial-gauging
representation), is bounded over the whole two-dimensional
subspace defined by $\left({SL(2,{\bf R})/ U(1)_{\rm
axial}}\right)_{k+4}$, and has its maximal (finite) value at the
horizon, i.e. the position of the T-dual NS5-branes. It would be
interesting to further investigate these issues, because one could
analyze various properties of the T-dual NS5-branes, as well as
their gravitational back-reaction by using the powerful
conformal-theory techniques developed in closed-string
perturbation theory.
All the solutions proposed so far to the infinite-coupling problem of the
five-brane background are based on dualities. As a consequence, there is
always a region of space--time where the coupling diverges. We will now show that it
is possible, instead, to modify this background in a way that (\romannumeral1)
the coupling remains finite everywhere and (\romannumeral2) that the string is
still described in terms of an exact superconformal theory.

\boldmath
\section{Interpolating between flat space and
three-sphere plus
linear dilaton}\label{match}
\unboldmath

The strong-coupling singularity that spoils the string perturbative
expansion
in the above five-brane background occurs at $r \to 0$ ($z\to -\infty$),
for the Neveu--Schwarz
branes. We will now propose a solution to this problem, which is inspired
from an electrostatic analogue. The case of D5-branes, where the divergence of
the coupling ($h_0 = 0$) occurs at $r \to \infty$ ($z\to + \infty$), cannot
be treated in the same way. Alternative solutions will be proposed later.

The divergence of the ordinary Coulomb field can be avoided by assuming a
spherically symmetric distribution of charge over a two-sphere centered at
the original point-like charge. We can similarly introduce a distribution of
five-branes over the transverse three-sphere, at some finite radius, say
$r=R$.  This amounts in adding to the bulk action (\ref{act}) a source
term of the form:
\begin{equation}
S_{\rm five-brane}= -\frac{N\, T_5}{2\pi^2} \int d^{10}x \,
 \sin^2\!\theta \,
\sin \varphi \, \delta(r-R) \left( \e^{\alpha
\gamma\,\phi_\gamma}\sqrt{-\hat{g}^{(6)}}+\tilde C_{6}\right),
\label{source}
\end{equation}
where $\tilde C_6$ is the dual of the two-index antisymmetric tensor.
Several remarks are in order here. In writing (\ref{source}),
we have chosen a gauge
in which $(t,\vec{x})$ are the world-volume coordinates of the five-branes.
Thus, the induced metric $\hat{g}^{(6)}_{\; \; ij}$ is just the
reduction of the
background metric $g^{(10)}_{\; \; \mu \nu}$
($\mu,\nu,\ldots \in 0,1,\ldots, 9$ and $i,j,\ldots \in 0,1,\ldots, 5$).
All five-branes are sitting at $r=R$
($z=Z$), and are homogeneously distributed over the $S^3$. Their density is
normalized so that the net number of five-branes be $N$. The dimension
$\alpha$ determines the coupling of the branes under consideration, and
consequently
their nature: D, NS or even more exotic extended objects. It is a free
parameter, which will be determined later.

The energy--momentum tensor of the source term (\ref{source}) is
\ba T^{\mu
\nu}_{\rm five-brane}(x)&=&\frac{2}{\sqrt{-g^{(10)}}}
\frac{\delta S_{\rm five-brane}}{\delta g^{(10)}_{\; \; \mu \nu}(x)}\nn \\
&=&-\frac{N\, T_5}{2\pi^2}\sin^2\!\theta \, \sin \varphi \,
\delta(r-R)\e^{\alpha \gamma \phi_\gamma}\, \delta^{\mu}_i \, \delta^{\nu}_j
\,
\hat{g}^{(6)\, ij} \sqrt{\frac{\hat{g}^{(6)}}{g^{(10)}}}. \ea This enables
us
to write the full equations of motion resulting from action (\ref{act}) plus
(\ref{source}). One can solve them by introducing the same ansatz as before
(Eqs. (\ref{5met})--(\ref{3form})). Compatibility now demands that
\begin{equation}
\alpha = - \frac{1}{2}.
\nn
\end{equation}
Hence, expressed in the sigma-model frame, Eq. (\ref{source}) exhibits the
following dilaton coupling: $\exp -\frac{3+\gamma}{2}\phi_\gamma$. For
$\gamma = + 1$ this is indeed the coupling of an NS5-brane,
while for $\gamma = - 1$
we recover the D5-brane. In either situation, the dilaton $\phi_{\gamma}(z)$
and the function $f(z)$ are given by (\ref{deq}) and (\ref{feq}), respectively,
while $h(z)$ now solves
\begin{equation}
\alpha' h(z)^{\frac{3}{4}}\e^{4z}\,\mysquare_{} h= -2 N \delta(z-Z).
\label{boxdelta}
\end{equation}
In writing the latter, we have expressed\footnote{The five-brane tensions
are
$T_{\; 5}^{\rm NS}=\frac{2\pi^2 \alpha'}{\kappa_{10}^2}$ and $T_{\, 5}^{\rm
D}=\frac{1}{4 \pi^{3/2} \kappa_{10} \alpha'}$ \cite{bachas}. They turn out
to be equal, once $\kappa_{10}$ is expressed in terms of $\alpha'$.}
$\kappa_{10}$ and $T_5$ in terms of $\alpha'$. The result is independent
of the nature of the brane.

\begin{figure}[htb]
\begin{center}
\epsfig{figure=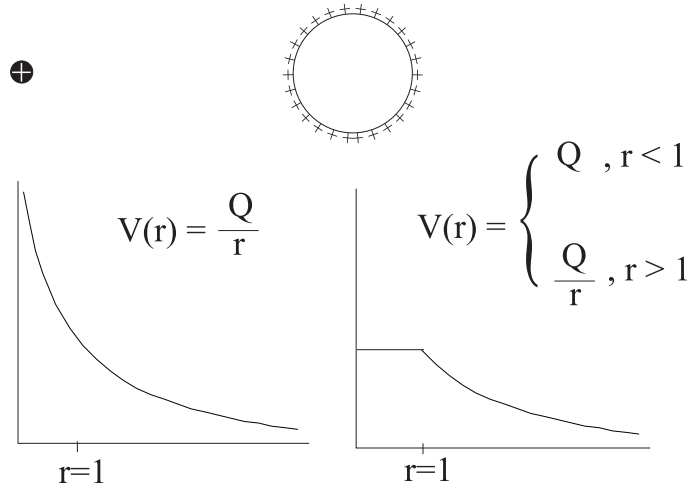, width=0.6\linewidth} \caption{\label{ea}
\small
\small Electrostatic analogue of the patched flat-space--NS5-brane solution.
The Coulomb potential of a point-like charge Q has a singularity at the
origin, which is resolved if the same
charge is distributed over the surface of a sphere (chosen here of
unit radius).
The Coulomb potential plays the role of the dilaton field and the
charge is $N$.}
\end{center}
\end{figure}

Replacing a point-like charge with a spherical distribution leads to the
same configuration outside the two-sphere, while the electric field  vanishes
inside (Gauss's law), avoiding thereby the Coulomb divergence.
This is depicted in
Fig.  \ref{ea}. The simplest solution to Eq. (\ref{boxdelta}),
where we set for simplicity $Z=0$ ($R=1$), is precisely an
analogue of that electrostatic example, as we have advertised previously:
\begin{equation}
h(z)=h_0 + N\e^{-(z+|z|)},
\label{electra}
\end{equation}
and, by using (\ref{feq}),
\begin{equation}
f(z)= N\, \Theta(z).
\nn
\end{equation}
For $r>1$ ($z>0$) we recover (\ref{heq}), while for $0<r<1$ ($z<0$) the
space
is flat since $h = h_0 +N$. Moving the brane sources from $r=0$ to a uniform
$S^3$ distribution at $r=1$ amounts therefore in excising a ball which contains
the
would-be near-horizon geometry, and replacing it with a piece of flat space.
The price to pay for this matching is the introduction of sources
uniformly distributed over $S^3$
and localized at $z=0$.

\begin{figure}[htb]
\begin{center}
\epsfig{figure=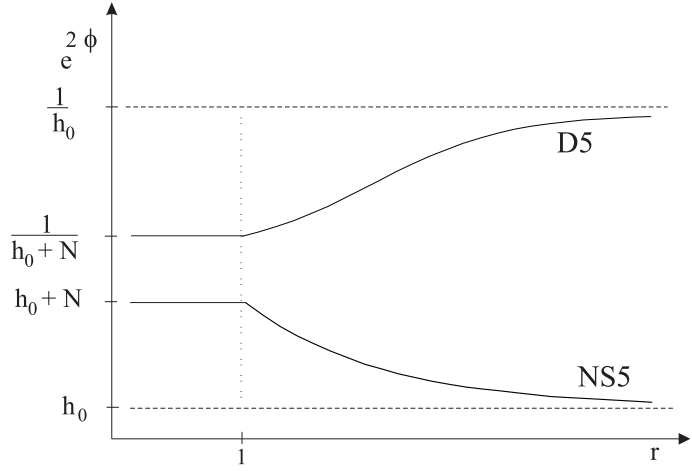, width=0.6\linewidth} \caption{\label{coupling}
\small
\small The string coupling of solution (\ref{electra}) is finite everywhere
for
the NS5-branes ($\gamma = + 1$) and its electrostatic analogue is given in
Fig. (\ref{ea}). However, it diverges when the sources are D5-branes
and $h_0 = 0$.}
\end{center}
\end{figure}

Concerning the dilaton field, Neveu--Schwarz and Dirichlet sources lead to
different pictures, according to Eqs. (\ref{deq}) and (\ref{electra}). For
NS5-branes, the excised ball removes altogether the divergent-coupling
region of the canonical neutral five-brane, and replaces it
with a constant one,
$g_{\rm s}^2 = h_0^{\vphantom 2} + N^{\vphantom 2}$. In the case of
D5-branes, the coupling inside the ball ($r<1$) becomes also constant,
$g_{\rm s}^2 = \left( h_0^{\vphantom 2} + N^{\vphantom 2}\right)^{-1}$.
These results are summarized in Fig. \ref{coupling}.

As we have already stressed, a remarkable situation is provided by $h_0 =
0$.
For negative $z$, the transverse space is flat, as for generic $h_0$. For
positive $z$, the geometry is that of a three-sphere of radius
$L=\sqrt{\alpha'
N}$ plus linear dilaton (see Eqs. (\ref{lidi}) and (\ref{nhg})). Both
patches
are type-II string backgrounds described in terms of exact $N=4$
superconformal
theories. In the case of Neveu--Schwarz sources, these are  free of
strong-coupling singularities.

\boldmath
\section{Orbifold and orientifold planes}\label{D5o}
\unboldmath

In fact, solution (\ref{electra}) is the only one that corresponds to a
NS5-brane distribution, with a string coupling that remains finite
everywhere.
However, this solution {\it fails to regularize the D5-brane configuration}
when the latter is singular, namely for $h_0 = 0$. Indeed, the string
coupling
is then divergent for large $z$, which is outside the excised ball. To
ensure
the finiteness of the coupling in the case at hand, we should instead
consider
a would-be dual solution:
\begin{equation}
h(z)=h_0 + N \e^{-z+|z|},
\label{ma}
\end{equation}
and, by using (\ref{feq}),
\begin{equation}
f(z)= N\, \Theta(- z).
\label{fma}
\end{equation}
Flat space would be the geometry outside the three-sphere at $r=1$, whereas
inside we would recover (\ref{heq}), that corresponds, for $h_0 = 0$, to the
three-sphere plus linear-dilaton. Here, the coupling constant would be {\it
finite everywhere for $\gamma = - 1$} (D5), while it would diverge at the
origin ($z \to -\infty$) for $\gamma = + 1$ (NS5).

The problem with (\ref{ma}) is that it does not solve Eq. (\ref{boxdelta}),
but solves a similar equation, with no negative sign (and $Z = 0$).
Such an equation can only
be obtained by assuming $T_5 <0$. We must therefore interpret solution
(\ref{ma}) as resulting from $N$ remote branes
($z \to -\infty $ i.e. $r =0$),
together with $N$ negative-tension objects localized at $z = 0$ ($r = 1$).
The net effect of the latter is to screen the charge sitting at $r=0$ so as to
ensure flat space for $r>1$. Again, this is the analogue of an electrostatic
configuration, where a point-like charge is surrounded by a homogeneous
spherical shell of opposite charge: outside the shell, the potential is
constant whereas it is Coulomb inside.

The negative-tension objects under consideration are of two kinds:
orientifold
planes if they are associated with D-branes ($\gamma = - 1$) or orbifold
planes
when they correspond to NS5-branes ($\gamma = + 1$). They cannot have
fluctuations in a unitary theory because the corresponding modes would be
negative-norm.

\boldmath
\section{Brane chains}\label{Ori}
\unboldmath

At this stage of the paper, it has become clear that consistent string
backgrounds can be constructed by using either five-branes or negative-tension
objects. The dilaton depends on $\gamma $, but the geometry, which is
governed by the function $h(z)$, does not (see Eqs. (\ref{5met}) and
(\ref{aeq})--(\ref{deq})). The function $h(z)$ depends, in turn, on whether
the source is a brane or an orbifold/orientifold plane (for simplicity, we will
call them generically ``orbifold planes", as long as we do not discriminate
Neveu--Schwarz and Dirichlet, i.e. as long as we do not discuss the issue of
the coupling but deal with the geometry only). The canonical solution
(\ref{heq}) corresponds to a source made of $N$ branes pushed at $z \to -
\infty$. In (\ref{electra}) those branes are at $z=0$, while solution
(\ref{ma}) is generated by $N$ branes at $z \to - \infty$ together with $N$
orbifold planes located at $z=0$. One might wonder what would the solution
look
like in the case where both five-branes and five-orbifold planes are
homogeneously
distributed over the $S_3$ and localized at certain discrete values of the
transverse coordinate $z$. In particular, one might also investigate the
conditions under which the corresponding geometry is the target space of an
exactly conformal sigma model. The aim of the present section is to clarify
these issues.

The generalization of Eq. (\ref{boxdelta}) for a network of sources reads
(we have used the explicit expression for the d'Alembert operator):
\begin{equation}
h'' + 2 h' = - 2 \sum_{k=1}^M N_{k}\, \lambda_k \e^{-2z_k}\,
\delta\left(z - z_k \right).
\label{netdel}
\end{equation}
It describes the geometry generated by $N_{k}>0$ objects (five-branes or
five-orbifold planes, depending on whether $\lambda_k= 1$ or $-1$) located at
$z = z_k$ for $k = 1, \ldots, M$. One of the two integration constants of the
above equation is the number of branes sitting at
$z \to -\infty$, $N_{0}\geq 0$;
those cannot be orbifold planes, that would generate negative $h(z)$, at
least in the asymptotic region of negative $z$.

Between two consecutive stacks of branes, the solution of (\ref{netdel}) is
of the general type:
\begin{equation}
h(z)=h_k+ \tilde N_k\e^{-2z}, \ \ {\rm for}\ \ z_k\leq z \leq
z_{k+1}\ \ k=0,\ldots,M
\label{netsol}
\end{equation}
($z_0$ and $z_{M+1}$ are meant to be $-\infty$ and $+\infty$, respectively).

The ``slopes" $\tilde N_k$ can be determined by computing the
discontinuities
of $h'$ at the locations of the sources. Equation (\ref{netdel}) enables us
to write
\begin{equation}
\tilde N_k - \tilde N_{k-1} = \lambda_k\, N_k.
\label{Qrec}
\end{equation}
Put differently, $\tilde N_k$ is the integrated charge from $-\infty$ to
$z_k$
included:
\begin{equation*}
\tilde N_k = N_0 + \sum_{i=1}^k \lambda_i\, N_i.
\end{equation*}
Continuity of $h(z)$, on the other hand, allows for the determination of the
$h_k$'s:
\begin{equation}
h_{k-1} = h_{M} + \sum_{i=k}^M
\lambda_i\, N_i \e^{-2z_i} \ \ k=1,\ldots,M,
\label{hksol}
\end{equation}
where $h_M$ is the other integration constant.

The choice of the charges $N_k$ and of their positions $z_k$ is
not completely arbitrary if one demands positivity of $h(z)$. In
order to analyze this issue, we will focus on specific charge
distributions where $\tilde N_k\geq 0 \ \forall k$. Although this
requirement is natural for $k=0$ in order to avoid $h(z)<0$ for
negative enough $z$, we could in principle allow some negative
integrated charges $\tilde N_k $ provided their positions as well
as the integration constant $h_M$ are chosen in such a way that
$h(z)$ remains positive everywhere. Our aim, however, is not to
analyze the most general case, but situations which resemble
(\ref{heq}), (\ref{electra}) and (\ref{ma}), where the total
integrated charge is non-negative for any $z$. Moreover, as we
will see later, backgrounds that can be described in terms of
exact conformal theories turn out to belong to the class at hand.

Our claim is that if all localized charges $N_k$ are chosen in such a way
that
the integrated ones, $\tilde N_k$, are never negative, then $h(z)$ is
non-negative, provided $h_M$ be non-negative, without any restriction on the
positions $z_k$. Indeed, together with $\tilde N_M\geq 0$, $h_M\geq 0$
guarantees that $h(z)\geq 0$ when $z\gg z_M$. On the other hand, according
to
Eq. (\ref{netsol}) and since $\tilde N_k\geq 0 \ \forall k$, $h(z)$ is
monotonically decreasing. So $h(z)$ is never negative.

The background described in Eq. (\ref{netsol}) is not expected to be an
exactly
conformal model for generic values of the data $N_{k}$ and $z_k$. For
every
$k=0,\ldots, M$, a necessary condition is that either $h_k$ or $\tilde N_k$
vanishes. In the latter case, the space is flat, whereas in the former it
contains a three-sphere plus a linear dilaton.

The starting point of our analysis is the recursion (continuity) relation
\begin{equation}
h_{k-1} = h_{k}+ \lambda_k\,
N_k\e^{-2z_k}
\nn
\end{equation}
(this has led to (\ref{hksol})). By construction, $N_k>0$ for $k=1,\ldots,
M$, while $N_0\geq 0$. Therefore, $h_{k}=0$ implies that $h_{k-1}=
\lambda_k\, N_k \exp\left(-2z_k\right) \neq 0$. This result
shows
that it is impossible to have two consecutive domains in $z$, both with
linear
dilaton and three-sphere, separated by a distribution of five-branes. The
only
allowed pattern, compatible with exact CFT\footnote{Strictly
speaking the $S^3$-plus-linear-dilaton background is an exact CFT only
when the antisymmetric tensor is of NS type.}, is an
alternation of flat-space and three-sphere-plus-linear-dilaton patches.

\begin{figure}[htb]
\begin{center}
\epsfig{figure=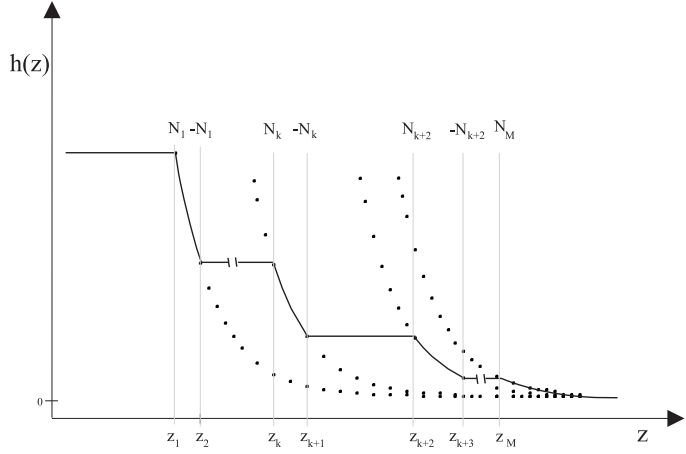, width=0.6\linewidth} \caption{\label{hcft} \small
\small A partial sequence of five-branes (at $z_{k}, z_{k\pm 2}, \ldots$)
and
five-orbifold planes (at $z_{k\pm 1}, z_{k\pm 3},\ldots  $) appears in the
center of this figure. The charges $\cdots < N_{k-2}<N_{k} < N_{k+2}<\cdots
$
are independent data, together with $z_{k}, z_{k\pm 2},\ldots $ The whole
drawing is the solution (\ref{hQinf}).}
\end{center}
\end{figure}

Hence, assuming that $h_k$ vanishes, we must impose  $\tilde N_{k \pm 1}=0$,
$h_{k \pm 2}=0$, etc. In particular, Eq. (\ref{Qrec}) now reads:
\begin{equation}
\tilde N_k = \lambda_k\, N_k = - \lambda_{k+1}\,
N_{k+1}.
 \label{Qreccft}
\end{equation}
Under these circumstances, a necessary and sufficient condition for $h(z)$
to
be non-negative is that $\tilde N_k>0$. This guarantees that $h(z)>0$
(\romannumeral1) for $z_{k-1} \leq z \leq z_k$, where $h(z) = h_{k-1}=
\tilde
N_k\exp\left(-2z_k\right)$; (\romannumeral2) for $z_k \leq z \leq z_{k+1}$,
where $h(z) = \tilde N_k \exp(-2z)$. From Eq. (\ref{Qreccft}), we therefore
learn
that $\lambda_k=1=-\lambda_{k+1}$ and $N_k = N_{k+1}$. Suppose that $N_k$
five-branes are sitting at $z_k$ with a total integrated charge $\tilde N_k =
N_k$. The function $h(z)$ is exponentially decreasing up to $z_{k+1}$, where
$N_{k+1} = N_{k}$ five-orbifold planes are localized. The total integrated
charge
$\tilde N_{k+1}$ vanishes again, and $h(z)$ remains constant and equal to
$h_{k+1}= N_{k+2}\exp\left(-2z_{k+2}\right)$ until $z_{k+2}$.
Another stack of $N_{k+2} > N_{k} $ five-branes appear at that point and the
process wraps back. Figure \ref{hcft} depicts the situation.

At this stage, it is important to notice that not all $N_k$'s and $z_k$'s
are
independent parameters. We have already observed that only the charges of,
say,
the five-branes (i.e. those with $\lambda_k = 1$) can be chosen arbitrarily;
the
charges of five-orbifold planes are then automatically determined. Moreover,
given
the positions of the five-branes, $z_{k}, z_{k\pm 2},\ldots$, we determine
those
of the five-orbifold planes:
\begin{equation}
z_{k-1} = z_k  - \frac{1}{2} \log \frac{N_k}{N_{k-2}},
\label{poscft}
\end{equation}
This relation shows in particular that the two sets of parameters, the
charges
$\{N_{k}$, $N_{k\pm 2}$, $\ldots  \}$ and the brane positions $\{ z_{k}, z_{k\pm
2},\ldots   \}$, though independent, must obey the following conditions:
\begin{equation}
z_{k-2} < z_{k} - \frac{1}{2} \log \frac{N_{k}}{N_{k-2}}.
\label{cftcon}
\end{equation}

For practical purposes (see Section \ref{loc}) it is useful to present the
solution $h(z)$ corresponding to a set of data (charges and positions) in
explicit and closed form. Let us suppose for concreteness that there is a no
charge at $z\to -\infty$: $N_{0} =  0$. Assuming an odd number of sources,
$M=2L - 1$, the independent data are chosen to be the charges of the
five-branes $\{N_{1}, N_{3}, \ldots, N_{2L-1}\}$, with $\lambda_{2s + 1}
=1$,
together with their positions $\{z_1, z_3, \ldots, z_{2L-1}\}$. The charges
$N_{2s}$ and positions $z_{2s}$ of the five-orbifold planes ($\lambda_{2s}
=-
1$) are given respectively by Eqs. (\ref{Qreccft}) and (\ref{poscft}) with
$k=2s + 1$. Assuming that inequality (\ref{cftcon}) is fulfilled, $h(z)$
reads:
\begin{equation}\begin{split}
h(z) =& \sqrt{N_{1} \, N_{2L-1}} \exp\left(-(z+z_1) -\left\vert z-z_1\right
\vert\right)\times\\
&\times\prod_{s = 1}^{L-1} \exp\left(\left\vert z - z_{2s +1} +\frac{1}{2} \log
\frac{N_{2s+1}}{N_{2s-1}} \right\vert - \left\vert z - z_{2s+1}\right
\vert\right).
\label{hQinf}
\end{split}
\end{equation}
We are in the case where $h(z)$ does not diverge at $z\to -\infty$, and
vanishes at  $z\to +\infty$: $h(z\leq z_1) = h_0 = N_{1} \exp(-2 z_1)$ and
$\lim_{z \to +\infty} h(z) = 0$. The transverse space is flat for  $z\leq
z_{1}$ and has the geometry of a three-sphere plus linear dilaton for $z\geq
z_{2L-1}$. This is a consequence of the absence of remote five-branes at
$z\to -
\infty$, and of the presence of $N_{2L-1}$ five-branes as last source. Figure
\ref{hcft} summarizes those features.

Any other situation can be obtained directly from Eq. (\ref{hQinf}), by
considering appropriate limits. In the limit $z_1 \to -\infty$, $N_{1}$
five-branes are pushed far away. The function $h(z)$ now diverges at $z \to
-\infty$: for $z\leq z_2$ it describes a three-sphere plus linear dilaton.
On
the other hand, when $z_{2L-1}$ is sent to $+ \infty$, the last sources are
five-orbifold planes localized at $z=z_{2L-2}$. From this point $h(z)$ will be
constant and the transverse space flat. Finally, both limits can be
simultaneously taken, so that for $z\leq z_2$  and $z\geq z_{2L-2}$, the
space
becomes, respectively, a three-sphere plus a linear dilaton, and flat. This
exhausts all the possibilities for constructing string backgrounds generated
by
five-branes and five-orbifold planes, that can be described in terms of
exact
CFT's. All these constructions have natural electrostatic
analogues, which consist of $M$ homocentric thin shells with charges
alternating in sign, with or without a charge at the origin ($r=0$ i.e.
$z\to
-\infty$).

As far as the string coupling is concerned (see Eq. (\ref{deq})), all
possible
situations appear: it might vanish, remain finite, or become infinite in one
or
both regions $r=0$ and $r\to \infty$, for either Dirichlet or Neveu--Schwarz
five-brane backgrounds.

\boldmath
\section{D5-NS5 transition in type-IIB theory}\label{dns}
\unboldmath

We have by now become familiar with the construction of elaborate
configurations of stacks of branes uniformly distributed over homocentric
three-spheres. As already stressed previously, the geometry (i.e. the
metric) generated by such configurations does not depend on the nature of
the five-branes/five-planes which are present. The dilaton, however, does:
the string coupling, $\exp 2\phi$, is equal to $h(r)$ or $h(r)^{-1}$ for
NS--NS or R--R backgrounds, respectively (see Eqs. (\ref{5met}) and
(\ref{aeq})--(\ref{deq})). If we try therefore to repeat the analysis of
Section 5 for type-IIB backgrounds whit both NS5-branes and D5-branes
(together with their negative-tension counterparts), we will
generically face discontinuities in the dilaton field at the location of
the source-shells. Alternatively, continuity requirement for the dilaton
leads to discontinuities in $h$, which amount to $\delta$-function
terms in the three-form field strength.

There is, however, one instance, where a continuous interpolation between
NS5-brane and D5-brane backgrounds is possible. In the type-IIB theory,
D5- and NS5-branes are S-dual. If two distinct regions of space--time,
hosting NS--NS and R--R backgrounds respectively, are
due to be smoothly patched together, this must happen at the S-self-dual point.

Let us be more concrete and consider $N$ D5-branes sitting at $r=0$.
For $0<r\leq r_{\rm S}$ those
create a three-sphere (transverse) geometry plus (finite) linear
dilaton and R--R antisymmetric tensor (Eqs. (\ref{deq}) and (\ref{heq})
with $\gamma = - 1$ and $h_0 = 0$). At radius $r=r_{\rm S}$, we introduce a set of
$N$ orientifold five-planes, uniformly distributed,
together with $N$ NS5-branes -- put differently, a O--NS bound-state.
The R--R charge is therefore screened,
so that the R--R antisymmetric tensor vanishes for $r\geq r_{\rm S}$,
while a NS--NS one is switched on.

From the geometry point of view, the presence of the O--NS bound-state
distribution is transparent:
\begin{equation}
h(r)=\frac{N}{r^2},
\nn
\end{equation}
everywhere, which ensures the $S^3$ factor. As already stressed, this
distribution alters the dilaton field:
\begin{equation}
\e^{2\phi}=\left\{ \begin{array}{lll}
\displaystyle {r^2\over N},&\phantom{aa} &0<r\leq r_{\rm S},\\ \\
\frac{\displaystyle N}{\displaystyle r^2},&\phantom{aa}&r_{\rm S}\leq r.
\end{array}\right.
\nn
\end{equation}
Continuity of the latter\footnote{Its first derivative is discontinuous,
though.} demands the parameter $r_{\rm S}$ be the S-self-dual point, namely
\begin{equation}
h(r_{\rm S})=\frac{1}{h(r_{\rm S})}.
\nn
\end{equation}
This implies $r_{\rm S}=\sqrt{N}$.

The antisymmetric tensors are discontinuous at $r=\sqrt{N}$. Inside we
have the R--R background of the D5-brane which is zero outside, and vice versa
for the NS--NS background. Notice that the source action to be added to
the bulk action is in this case
\begin{equation}\begin{split}
S_{\rm O-NS}=& \frac{N\, T_5}{2\pi^2} \int d^{10}x \, \sin^2\!\theta \, \sin
\varphi \, \delta(r-\sqrt{N}) \left( \e^{{1\over 2}
\phi}\sqrt{-\hat{g}^{(6)}}+\tilde B_{6}\right)
\\
-&\frac{N\, T_5}{2\pi^2} \int d^{10}x \, \sin^2\!\theta \, \sin \varphi \,
\delta(r-\sqrt{N}) \left( \e^{-{1\over 2}\phi}\sqrt{-\hat{g}^{(6)}}+\tilde
C_{6}\right).
\nn
\end{split}\end{equation}
In the above configuration, the coupling is small everywhere except
near $r=\sqrt{N}$, where it is of order one. This is the price to pay for
continuously interpolating between NS--NS and R--R backgrounds. We could
excise the order-one-coupling region, by separating the O5-plane and
NS5-brane shells. This amounts, however, to abandon the continuity
requirement. We will not pursue any further this issue.

\boldmath
\section{Gravitational fluctuations}\label{loc}
\unboldmath

In this section, we consider gravitational fluctuations of the
five-brane solutions derived in Sections \ref{D5o} and \ref{Ori}.
Our motivation is to analyze the low-lying spectrum of states,
which can be eventually compared with exact CFT results, available
in the present setting \cite{KK}. Furthermore, this analysis is
important for clarifying the issue of localization of the states
in the vicinity of the branes. Our conclusion is that in the
framework of the five-brane and five-orbifold backgrounds
presented so far, gravitons and their K--K descendants
have a mass gap in accordance with the CFT analysis,
\cite{kounnas,K93,AFK,KK}. Other fluctuations (Neveu--Schwarz,
Ramond--Ramond, \dots) can be studied in the same manner, but this
is beyond the scope of the present work.

We will restrict ourselves to gravitational fluctuations that are
longitudinal
to the brane. To this end we consider small perturbations of the background
metric (\ref{5met}), (\ref{aeq}) and (\ref{beq}), of the form
\begin{equation}
ds^2= \alpha'\,
h(r)^{-\frac{1}{4}}\left(\eta_{ij}+\rho_{ij}\right)\,dx^i\,dx^j +
\alpha' \, h(r)^{\frac{3}{4}}\left(dr^2+r^2 d\Omega_3^2\right),
\label{perturbation}
\end{equation}
where\footnote{Indices $i,j,\ldots$ are
raised with the flat metric $\eta_{ij}$.}
$\{x^i, i=0, \ldots,5\}=\{t,\vec{x}\}=x$. The linearized
Einstein equations in the transverse, traceless gauge,
$\rho^i_{\, i}= \partial^i\rho_{ij}=0$,
taking also into account the sources,
reduce to the covariant scalar equation \cite{BS,Csaki:2000fc}:
\begin{equation}
\mysquare_{} \rho_{ij}=0,
\label{peq}
\end{equation}
where the d'Alembert operator is that of the unperturbed metric.
The solutions
of the above equation belong to the gravitational K--K sector.
Considering a K--K mode with mass $M$ ($M^2$ is measured in units
of $1/ \alpha'$), and assuming the factorization
$\rho_{ij}\left(x,r,\Omega\right)=
\tilde\rho_{ij}\left(x\right)\,
\phi(r,\Omega)$,
with
$\mysquare_6\tilde\rho_{km}\equiv
\eta^{ij} \partial_i \partial_j \tilde\rho_{km}=M^2
\tilde\rho_{km}$,
Eq. (\ref{peq}) reduces to its transverse part
\begin{equation}
\left(\frac{1}{r^3}\partial_r r^3\partial_r +
\frac{1}{r^2}\triangle_{S^3}+
M^2\, h(r)\right)\phi(r,\Omega) =0,
\label{peqKK}
\end{equation}
where $\Delta_{S^3}$ is the
Laplacian operator on the three-sphere.
We can  further decompose the
transverse-space dependence of the fluctuations:
$\phi(r,\Omega)=r^{-3/2}\,y_\ell(r)\,D^\ell_{nn'}(\Omega)$,
where $D^\ell_{nn'}(\Omega)$ form a complete set of orthonormal functions on
$S^3$:
\begin{equation}
\triangle_{S^3} D^\ell_{nn'}(\Omega) = -\ell (\ell + 2)
D^\ell_{nn'}(\Omega).
\nn
\end{equation}
Then the radial equation reads: \ba
\left[-\frac{d^2}{dr^2}+\frac{1}{r^2}\left(\ell+\frac{1}{2}\right)
\left(\ell+\frac{3}{2}\right)\right]\,y_\ell(r)=M^2\, h(r)\,y_\ell(r),
\label{SL} \ea which is a the Sturm--Liouville equation.

The natural inner product for the radial wave functions $y_\ell(r)$ is
obtained
by analyzing the normalization of the kinetic terms for $\rho_{ij}$ as they
appear when the fluctuations $g_{ij} = \eta_{ij}+\rho_{ij}$ are introduced
in
the action (Eqs. (\ref{act})  and (\ref{source})), and the latter is
expanded.
We obtain:
\begin{equation}
\parallel y_\ell \parallel^2 =
\int_0^\infty dr \, h(r) \, y_\ell(r)^2.
\label{inn}
\end{equation}
With this precise inner product, the Sturm--Liouville operator (in the
square
brackets of the lhs of (\ref{SL})) is self-adjoint, provided some
appropriate
boundary conditions are imposed, which include those we will consider here:
$y_\ell (0)=0$ and $\lim_{r\to \infty}y_\ell^\prime (r)=0$. This property
ensures the existence of a complete set of orthonormal eigenfunctions, whose
$M^2$-spectrum is real, non-degenerate, bounded from bellow, and contains at
least a continuous part.

In order to determine the spectrum we need a specific background. We will
consider for simplicity the single five-brane shell solution (\ref{electra})
with $h_0 = 0$:
\begin{equation}
h(r)=\left\{ \begin{array}{ll}
\displaystyle {N\over R^2}, &0\leq r \leq R,\\
\\
\frac{\displaystyle N}{\displaystyle r^2},&R\leq r.
\end{array}\right.
\label{hsing}
\end{equation}
The eigenfunctions of (\ref{SL}) are obtained by following the standard
strategy. We first solve Eq. (\ref{SL}) for $0\leq r \leq R$ and keep only
solutions that satisfy $y_\ell (0)=0$. For $M^2\neq 0$, those are Bessel
functions
\begin{equation}
y_\ell (r) = A_\ell \, \sqrt{r} \, J_{\ell + 1}
\left({M\sqrt{N}r \over R} \right)
, \ \ 0\leq r \leq R,
\label{0r1}
\end{equation}
behaving like $r^{\ell + 3/2}$ in the vicinity of $r=0$; $A_\ell$ is an
arbitrary, real constant. This solution holds even for $M^2 < 0$ (in that
case
$A_\ell$ is a real number times $i^{-(\ell +1)}$). Notice that
$\sqrt{r}N_{\ell
+ 1}\left(M\sqrt{N}r/ R \right)$ is also a solution, which must be
discarded because of its bad behavior ($r^{-\ell - 1/2}$) at the origin. For
$M^2 = 0$, the only acceptable solution is
\begin{equation}
y_\ell (r) = A_\ell\, \sqrt{r} \left(  {r\over R}\right)^{\ell + 1}
, \ \ 0\leq r \leq R , \ \ M^2 = 0.
\nn
\end{equation}

We must now solve Eq. (\ref{SL}) for $r\geq R$. Depending on $\ell$ and $M$,
the behavior is either oscillatory or power-law:
\begin{equation}
y_\ell (r) = B_\ell \, \sqrt{r} \, \cos \left( \sqrt{M^2 N -(\ell + 1)^2}
\log
\left({r \over R}\right)+ \Phi_\ell \right)\ , \ \
M^2 \geq \frac {(\ell + 1)^2}{N},
\label{1r}
\end{equation}
where $B_\ell$ and $\Phi_\ell$ are arbitrary, real constants, and
\begin{equation}
y_\ell^{\vphantom +} (r) = \sqrt{r}
\left(C_\ell^-
\left(\frac{R}{r}\right)^{\left((\ell + 1)^2 - M^2 N\right)^{1/2}} +
C_\ell^+\left(\frac{r}{R}\right)^{\left((\ell + 1)^2 - M^2 N\right)^{1/2}}
\right) \ , \ \ M^2 \leq \frac {(\ell + 1)^2}{N}.
\label{1rp}
\end{equation}
The latter holds even for $M^2 \leq 0$, and $C_\ell^\pm$ are again real and
arbitrary. In fact, we must set $C_\ell^+$ to zero, because with the inner
product (\ref{inn}) and (\ref{hsing}), $r^{1/2 + \sqrt{(\ell + 1)^2 - M^2
N}}$
is not even $\delta$-function normalizable. However, $r^{1/2 - \sqrt{(\ell +
1)^2 - M^2 N}}$ is normalizable, while (\ref{1r}) is $\delta$-function
normalizable (both satisfy the previously advertised boundary condition at
infinity). The spectrum is therefore expected to be continuous for $M^2 N
\geq
(\ell + 1)^2$ and discrete otherwise.

The complete determination of the eigenfunctions is achieved by
requiring continuity at $r=R$. On one hand, continuity of $y_\ell$
fixes $B_\ell^{\vphantom -}$ or $C_\ell^-$ in terms of
$A_\ell^{\vphantom -}$, leaving only an overall free
normalization. Continuity of the logarithmic derivative, on the
other hand, allows for the computation of the phase $\Phi_\ell$ if
$M^2 \geq {(\ell + 1)^2}/{N}$ (continuous spectrum), and the
positions of the discrete mass levels if $M^2 \leq {(\ell + 1)^2}/
{N}$. We find
\begin{equation}
\Phi_\ell = - {\rm arctan\, } \left(\frac{1}{\sqrt{1-(\ell +1)^2/M^2 N}}
\frac{J'_{\ell +1}\left(M\sqrt{N}\right)} {J_{\ell
+1}\left(M\sqrt{N}\right)}\right)
\nn
\end{equation}
for the continuous spectrum, while {\it the discrete spectrum turns out to
be
empty:} no mass-squared levels (positive, zero or negative) exist for $M^2 <
{(\ell + 1)^2}/{N}$. We therefore conclude that there is a {\it mass gap}
$M_{\rm gap}= 1/\sqrt{N}$ (in units of $M_{\rm s}$) in agreement with CFT
\cite{kounnas,K93,AFK,
KK}. Choosing $B_\ell
=
\left( N-(\ell + 1)^2 / M^2 \right)^{-1/4}$, the corresponding (complete)
set
of eigenfunctions is normalized as
\begin{equation}
\left(y_\ell^{(1)}, y_\ell^{(2)}\right)=\frac{\pi}{2}\,
\delta\left(M^{(1)} -
M^{(2)}\right),
\nn
\end{equation}
according to the inner product (\ref{inn}). For different $\ell$'s,
orthogonality is guaranteed by the spherical functions $D^\ell_{nn'}$.

Our result deserves several comments. First, had we considered instead of
(\ref{hsing}) a more general conformal background of the type (\ref{hQinf}),
our conclusions would not have been modified. For $0\leq r \leq r_1\equiv
\exp
z_1$ we have indeed the solution (\ref{0r1}) with $N$ replaced with
$N_1/r_1^2$, while (\ref{1r}) and (\ref{1rp}) -- with $C_\ell^+ = 0 $ -- are
valid for $r\geq r_{2L-1}$ with $N_{2L-1}$ instead of $N$. Continuity
constraints for $y_\ell (r)$ and $y_\ell'(r)$ propagate through all
intermediate-brane positions, determine completely the intermediate
solutions,
and eventually the spectrum of longitudinal gravitational fluctuations: this
is
a continuous spectrum above a mass gap $M_{\rm gap}= 1/\sqrt{N_{2L-1}}$.

\def\ri{R_{\infty}}
\boldmath
\section{Holography}\label{ispec}
\unboldmath

The framework is here the NS5-brane of type-IIA string theory. The
non-normalizable modes in this background are expected to correspond, via the
holographic principle \cite{maldacena,Itzhaki:1998dd,ABKS}, to off-shell
operators of the decoupled world-volume theory of the five-brane. The
boundary
is at $r=R_{\infty}\to \infty$.

In the case under study, however, the identification of the
radial direction as a renormalization-group flow is less clear.
The space-time metric (in the string frame),
\begin{equation}
ds^2_{\sigma}= {\alpha'}\, dx^2_{\vphantom 3}+{\alpha'}\,
h(r)\left(dr^2_{\vphantom 3}+r^2_{\vphantom 3} d\Omega_3^2\right),
\nn
\end{equation}
is invariant under a rescaling of $r$. The only
effect is to shift appropriately the position of the shell. If the shell is
at
the origin the metric is strictly invariant. The string coupling on the
other
hand $g_{\rm s}^2=\e^{2\phi}=h(r)$ scales to zero as we approach the boundary.

The decoupling limit is \cite{Itzhaki:1998dd}
\begin{equation} g_{\rm s}=\e^{\phi}\to 0 \ , \ \
\ell_{\rm s}=\sqrt{\alpha'} \ \ {\rm and}\ \ U={r\over g_{\rm s}
\ell_{\rm s}} \ \ {\rm fixed,}
\nn
\end{equation}
where $U$ is the tension\footnote{Remember that the dimensionless radial
coordinate $r$ measures distances in units of $\ell_{\rm s}$.} of a
D2-brane stretched between NS5-branes. It is the tension of a
world-volume string  and corresponds to a Higgs vev of the LST
(boundary theory). Thus, the boundary theory has no dimensionless coupling
constant and its scale is set by $\ell_{\rm s}$. The dilaton is given by
$\e^{2\phi}={N\over U_{\vphantom s}^2\ell_{\rm s}^2}$, and vanishes
at infinity. Using the relation
between $r$ and energy ($U$), this implies that the region of small $r$
corresponds to the infra-red conformal theory.

The presence of the shell at $r=R$, modifies somewhat the picture above.
First we choose the position of the shell
(we will from now on rescale $r\to g_{\rm s} r$) so that $R\gg\sqrt{N}$.
Large $N$ implies
that all curvatures are small everywhere (so that stringy corrections can be
neglected). In particular, the supergravity description is good at length
scales larger than the string scale $\ell_{\rm s}$.
The condition on $R$ also implies that the string coupling in the bulk
is small everywhere. Thus, the supergravity description is reliable on
the whole space.

In the background under consideration, the world-volume theory undergoes a
Higgs phenomenon at an energy scale ${1\over \ell_{\rm s} R}$.
Since $R\gg\sqrt{N}$ this has modified the effective field theory at
lower scales. Below this scale, the coupling no longer runs.

There is another approximation that is relevant here, and this has to do
with the continuous distribution of five-branes on the shell. String theory
implies that the NS5-charge is quantized. Thus, if the source at $R$
is composed of $N$
NS5-branes, the distribution is quasi-continuous. In fact, the average
distance (in transverse space) between the $N$ five-branes distributed
over the sphere is
$L\simeq N^{-1/3} \ell_{\rm s} R \gg N^{1/6} \ell_{\rm s}$.
Thus, at large $N$, $L$ is much larger than the string scale but much
smaller that the characteristic scale $\sqrt{N} \ell_{\rm s}$ of the
world-volume theory.

We can summarize the previous discussion as follows:
the supergravity description with
$SO(4)$ symmetry is valid at length scales larger than $N^{1/6}\ell_{\rm s}$.
At length scales larger than $\sqrt{N}\ell_{\rm s}$
the LST can be replaced by an effective field theory.
At length scales smaller than $N^{1/6}\ell_{\rm s}$ but larger
than $\ell_{\rm s}$ the supergravity description is valid
but $SO(4)$ symmetry is broken.

We will now proceed to apply the holographic principle as implemented in
\cite{Witten:1998qj} and calculate
boundary correlators.
In particular, we will focus on the (descendants) of the six-dimensional
graviton, itself dual to the world-volume
stress tensor. We have shown that the six-dimensional graviton
($\ell=0$)-fluctuation and its K--K descendants ($\ell > 0$) satisfy
Eq. (\ref{peq}) that can be recast in the form (\ref{peqKK})
or (\ref{SL}).
For a given set of boundary data,
$\bar\rho_{ij}^{\ell nn'}(x,\Omega)=
\bar\rho_{ij}^{\ell}(x)\, D^\ell_{nn'}(\Omega)$,
the corresponding partial-wave bulk solution,
$\rho_{ij}^{\ell nn'}(x,r,\Omega)=
\rho_{ij}^{\ell}(x,r)\, D^\ell_{nn'}(\Omega)$,
can be expressed in terms of the bulk-to-boundary propagator
$G_{\ell n n'} (x,r,\Omega)= G_{\ell} (x,r)\, D_{nn'}^\ell (\Omega)$:
\begin{equation}
\rho_{ij}^{\ell}(x,r)=
\int d^6x'\, G_{\ell} (x-x',r)\,
\bar\rho_{ij}^{\ell}(x').
\label{conv}
\end{equation}

The radial part of the above propagator satisfies
\begin{equation}
\left(\frac{1}{r^3}\partial_r r^3\partial_r -
\frac{\ell (\ell +2)}{r^2}+
h(r)\,\mysquare_6 \right) G_\ell (x, r) =0
\label{btb}
\end{equation}
with the boundary condition
\begin{equation}
\lim_{r\to \ri}G_\ell (x,r)=\delta^{(6)}(x).
\nn
\end{equation}
We will work from now on in Euclidean space, and Fourier transform
the six-dimensional part
\begin{equation}
G_\ell (x,r)={1\over {\alpha'}^3}\int{d^6p\over (2\pi)^6}
\e^{i p \cdot x} \, G_\ell (p,r),
\nn
\end{equation}
so that Eq. (\ref{btb}) becomes:
\begin{equation}
\left(\frac{1}{r^3}\partial_r r^3\partial_r -
\frac{\ell (\ell +2)}{r^2}-
h(r)\,p^2 \right) G_\ell (p, r) =0
\label{btb4}
\end{equation}
with boundary condition
\begin{equation}
\lim_{r\to \ri}G_\ell (p,r)=1.
\label{btbc}
\end{equation}

The regular solution to the bulk equation (Eq. (\ref{btb4})) is
\begin{equation}
G_\ell(p,r)=\left\{ \begin{array}{ll}
\displaystyle{A_\ell(p)\over r}\, I_{\ell+1}
\left(\displaystyle{\sqrt{p^2 N}r\over R}\right),\hfill
0\leq r \leq &R,\\
\\
\displaystyle{C^-_\ell(p)\over r} \left(\displaystyle{R\over r}
\right)^{\sqrt{(\ell+1)^2+p^2 N}}
+\displaystyle{C^+_\ell(p)\over r} \left(\displaystyle{r\over R}
\right)^{\sqrt{(\ell+1)^2+p^2 N}},\ R \leq &r.
\end{array}\right.
\label{btb3}
\end{equation}
The boundary condition (\ref{btbc}) is satisfied provided
\begin{equation}
{C^+_\ell(p)\over \ri}
=\left( {R\over \ri}\right)^{\sqrt{(\ell+1)^2+p^2 N}}.
\nn
\end{equation}
The rest of the coefficients are determined from the usual matching
conditions (continuity of the propagator and its logarithmic derivative).
We obtain:
\begin{equation}
{A_\ell(p)\over \ri}=
\left({R\over \ri}\right)^{\sqrt{(\ell+1)^2+p^2 N}}
{2\sqrt{(\ell+1)^2+p^2 N}\over
\sqrt{(\ell+1)^2+p^2 N}\, I_{\ell+1}\left(\sqrt{p^2 N}\right)
+\sqrt{p^2 N}\, I'_{\ell+1}\left(\sqrt{p^2 N}\right)}
\nn
\end{equation}
and
\begin{equation}
{C^-_\ell(p)\over \ri} =
\left({R\over \ri}\right)^{\sqrt{(\ell+1)^2+p^2 N}}
{\sqrt{(\ell+1)^2+p^2 N} \,
  I_{\ell+1}\left(\sqrt{p^2 N}\right)-\sqrt{p^2 N}\,I'_{\ell+1}
\left(\sqrt{p^2 N}\right)\over
\sqrt{(\ell+1)^2+p^2 N}\, I_{\ell+1}\left(\sqrt{p^2 N}\right)
+\sqrt{p^2 N}\, I'_{\ell+1}\left(\sqrt{p^2 N}\right)}.
\nn
\end{equation}

In order to make contact with holography, we would like to analyze the
dynamics of the six-dimensional gravitational perturbations
$\rho_{ij}(x^\mu)$ considered in (\ref{perturbation}), from a slightly
different point of view. Let us write down the linearized action for those
fields, as it appears when expression (\ref{perturbation}) is plugged into
Eq. (\ref{act}). We obtain (in the transverse, traceless gauge considered
so far):
\begin{equation}
S_2={3 {\alpha'}^4\over
2\kappa_{10}^2}\int d^6x\, d^4y
\left(h\,
\p^i\rho_{jk}\, \p_i\rho^{jk}+\p^a\rho_{jk}\, \p_a\rho^{jk}\right),
\nn
\end{equation}
where $\{y^a, a=6,\ldots,9\}$ are transverse
coordinates such that
$dr^2+r^2 d\Omega_3^2=\sum_a (dy^a)^2$
with\footnote{Indices $a,b,\ldots$ are raised with the
flat metric $\delta_{ab}$, while $i,j,\ldots$ are with $\delta_{ij}$ (we
have trade $\eta_{ij}$ for its Euclidean counterpart).
The Laplacian operator associated with the -- euclideanized --
unperturbed metric
(Eqs. (\ref{5met})--(\ref{beq})) now reads:
$\alpha' \mysquare = h^{1/4} \mysquare_6 + h^{-3/4} \mysquare_4$,
where $\mysquare_4 = \delta^{ab}\p_a \p_b$ and
$\mysquare_6 = \delta^{ij}\p_i \p_j$.}
$\sum_a (y^a)^2 = r^2$.
The Gibbons--Hawking boundary term, the antisymmetric-tensor and dilaton
terms, as well as the source term vanish.
By using the partial-wave expansion of $\rho_{ij}(x^\mu)$, integration by
parts, ortho-normality relations for the $D^\ell_{nn'}(\Omega)$, equations
of motion, as well as
Eq. (\ref{conv}) and the above expressions for the bulk-to-boundary
propagator (see Eq. (\ref{btb3})), we can expand $S_2$ in partial waves,
$$
S_2^{\phantom \ell}= \sum_{\ell=0 }^{\infty} S_2^{\ell},
$$
and determine $S_2^{\ell}$ in terms of the
Fourier-transformed boundary data $\bar\rho^\ell_{ij}(p)$:
\begin{equation}
S_2^{\ell}= -{3\over 2\kappa_{10}^2{\alpha'}^2} \int d^6p\,
\bar\rho_{ij}^{\ell}(p) \, \bar\rho^{\ell ij}(-p)\, \ri^2\,
\Upsilon_\ell.
\nn
\end{equation}
We have introduced the function
\begin{equation}
\begin{split}
\Upsilon_\ell (p)=&
\left(1-\sqrt{(\ell+1)^2+p^2 N}\right)
+
2\chi_\ell\left({R\over\ri}\right)^{2\left((\ell+1)^2+p^2 N\right)^{1/2}} \\
&
+\chi_\ell^2 \left(1+\sqrt{(\ell+1)^2+p^2 N}\right)
\left({R\over\ri}\right)^{4\left((\ell+1)^2+p^2 N\right)^{1/2}},
\end{split}
\nn
\end{equation}
where $\chi_l$ also depends on $p^2 N$
\begin{equation}
\chi_\ell(p)={
\sqrt{(\ell+1)^2+p^2 N} \, I_{\ell+1}\left(\sqrt{p^2 N}\right)
-
\sqrt{p^2 N}\, I'_{\ell+1}\left(\sqrt{p^2 N}\right)
\over
\sqrt{(\ell+1)^2+p^2 N} \, I_{\ell+1}\left(\sqrt{p^2 N}\right)
+
\sqrt{p^2 N}\, I'_{\ell+1}\left(\sqrt{p^2 N}\right)
}
\nn
\end{equation}
with the following asymptotic behavior:
\begin{equation}
\chi_\ell(p)={2\ell^2+9 \ell + 11
\over 8(\ell+1)^2\, (\ell+2)\, (\ell+3)}
\, p^2 N+{\cal O}\left(p^4 N^2\right)\ , \ \
 Np^2\ll 1,
\nn
\end{equation}
while
\begin{equation}
\chi_\ell={1\over 4\sqrt{p^2 N}}+{\cal O}\left({1\over p^2 N}\right) \ , \ \
Np^2\gg (\ell+1)^2.
\nn
\end{equation}
Thus, $\chi_\ell$ vanishes at $p\to 0$ and $p \to \infty$, and has a
maximum at $\sqrt{p^2N}  \sim \ell+1$.

We should note that the three different terms have different asymptotic
behavior as the renormalization screen
moves to infinity ($\ri\to \infty$), with the first term giving the leading
contribution.
In the region of validity of the supergravity approximation, $p^2 N$ can
be either very small or very large,
(the first region corresponds
to the effective-field-theory region, while the second to the LST region),
so
$\sqrt{(\ell+1)^2+p^2 N}$ cannot be expanded in a sequence of local terms.
Thus, if we insist on keeping the stringy physics of LST, we must renormalize
$\bar\rho_{ij}^\ell \to {\alpha'}^{-1/2}\ri^{-1} \bar\rho_{ij}^\ell$, in
which case the terms
proportional to $\chi_\ell$ and $\chi_\ell ^2$ vanish in the
$\ri\to \infty$ limit with the result:
\begin{equation}
S_2^{\ell}= -{3\over 2\kappa_{10}^2 {\alpha'}^3}\int d^6 p\,
\bar\rho_{ij}^\ell(p)\, \bar\rho^{\ell ij}(-p)
\left(-1+\sqrt{(\ell+1)^2+p^2 N}\right).
\label{btb5}
\end{equation}

At this point we can take advantage of the transverse-tracelessness
conditions, which in momentum space read
$\rho_{i}^i=p^i \rho_{ij}=0$, to ``covariantize" the two-point correlator
appearing in Eq. (\ref{btb5}):
\begin{equation}
S_2^{\ell}=-{3\over 4\kappa_{10}^2 {\alpha'}^3}\int d^6p\,
\bar\rho_{ij}^\ell(p)\, \bar \rho_{km}^\ell(-p)\,
F_{\ell}^{ij;km}(p)
\nn
\end{equation}
with
\begin{equation}
F_{\ell}^{ij;km}(p)=\left(\pi^{ik}\, \pi^{jm}+\pi^{im}\, \pi^{jk}-
{2\over 5}\, \pi^{ij}\, \pi^{km}\right)\left(-1+\sqrt{(\ell+1)^2+p^2 N}\right),
\nn
\end{equation}
where
\begin{equation}
\pi^{ij}=\delta^{ij}-{p^i\, p^j \over p^2}
\nn
\end{equation}
are the projectors that impose
conservation of the boundary stress tensor  and its K--K descendants. Thus, we
expect that
\begin{equation}
\left\langle T_{\ell}^{ij}(p)\, T_{\ell'}^{km}(-p)\right\rangle =-{3\over
4\kappa_{10}^2 {\alpha'}^3}\, F_{\ell}^{ij;km}(p)\, \delta_{\ell \ell'},
\label{tt}
\end{equation}
which is compatible with
conservation and tracelessness of the stress tensor and its cousins. Notice
that the two-point function is insensitive to the presence of the shell.
This will no longer be true for higher correlators.

Fourier transforming  in configuration space we obtain:
\begin{equation}
\left\langle T_{\ell}^{ij}(x)\, T_{\ell}^{km}(0)\right\rangle
=\left(\hat\pi^{ik}\, \hat \pi^{jm} +
\hat \pi^{im}\, \hat \pi^{jk} - {2\over 5}\hat\pi^{ij}\, \hat\pi^{km}\right)
F_{\ell}(x)
\nn
\end{equation}
with
\begin{equation}
 F_{\ell}(x)= -{3\over 4\kappa_{10}^2  {\alpha'}^3}
\int d^6p \, \e^{ip\cdot x} \left(-1+\sqrt{(\ell+1)^2+p^2 N}\right)
\nn
\end{equation}
and
\begin{equation}
\hat \pi^{ij}=\delta^{ij}-{\p^i\, \p^j\over \mysquare_{6} }.
\nn
\end{equation}

It is instructive here to pause and calculate the canonical
dimension of
$T^{ij}_{\ell}(p)$. For this we need to remember that
$\bar \rho_{ij}^{\ell}(x)$ has
canonical mass dimension zero, and its Fourier transform
$\bar\rho_{ij}^{\ell}(p)$
mass dimension $-6$. For renormalization purposes we have absorbed a factor of
$\sqrt{\alpha'}\ri$ in it so its canonical dimension changed to $-7$.
Thus, the canonical mass
dimension of $T^{ij}_{\ell}(p)$ is $+7$. This is reflected in (\ref{tt}),
taking into account that $\kappa_{10}$ has dimension $-4$.

We can evaluate the Fourier transform by using the following formula:
\begin{equation}
\begin{split}
-1 + \sqrt{(\ell+1)^2+p^2 N}
=& {1\over \sqrt{\pi}}\int_{(\ell+1)^2+p^2N}^1
dz\,
\int_0^{\infty}du~\e^{-zu^2}\\
=& {1\over \sqrt{\pi}}\int_0^{\infty}{du\over u^2}
\left(\e^{-u^2\left((\ell+1)^2+p^2N\right)}-\e^{-u^2}\right).
\end{split}
\nn
\end{equation}
Assuming $|x|\not =0$, and expressing $\kappa_{10}$ and $\alpha'$ in terms
of $M_{\rm s}$, we obtain the result\footnote{Remember
that in our conventions masses,
momenta and coordinates are all dimensionless (i.e. measured in units of
$M_{\rm s} = 1/\sqrt{\alpha'}$).}:
\begin{eqnarray}
F_{\ell}(x)&=&
-{3\over  2(2\pi)^4}{M_{\rm s}^{14}\over |x|^7}\, \e^{-{(\ell+1)|x|\over
\sqrt{N}}}\times\nn\\
&& \times
\left(15+15\, {(\ell+1)|x|\over \sqrt{N} }+
6\left({(\ell+1)|x|\over \sqrt{N}}\right)^2
+\left({(\ell+1)|x|\over \sqrt{N}}\right)^3
\right).
\nn
\end{eqnarray}
Note that for the modes with
$\ell\not =0$, there is a contact term proportional to a
$\delta$-function at $x=0$.

The above result implies that at large
distances the effective mass is
$M_{\rm eff}={(\ell+1)\over \sqrt{N}}$. This is the same as the mass gap we
found for the normalizable modes (previous section).
We also observe that the ``stringy" two-point
function is independent of the presence of the shell.
This is not true for higher correlators because those
involve the bulk-to-bulk propagator (which is computed in the appendix)
and depend on the presence of the shell.
A direct calculation (up to projectors) gives:
\begin{eqnarray}
F_3(p_1,p_2,p_3)&\sim&\lim_{\ri\to \infty}\int_0^{\infty} dr\,
G_{\ell_1}(p_1,\ri)\,
G_{\ell}(p_2,r)\,
G_{\ell}(p_3,r)\,
B_{\ell_1}(p_2+p_3,\ri;r) \times\nn\\
&&\times\delta(p_1+p_2+p_3),\nn
\end{eqnarray}
and the result depends on the position of the shell
via the bulk-to-bulk propagator $B_{\ell}(p,\ri;r)$.

If, however, we restrict ourselves to the effective-field-theory region
$\sqrt{p^2N}\ll 1$, we can expand the square root
in a power series in $p^2N$, keeping the first few terms. Such terms can be
removed by local counterterms, and the renormalized boundary data become
\begin{equation}
\bar \rho^\ell_{{\rm ren},ij}(p)=
\left({\ri\over R}\right)^{-\ell-{p^2 N\over 2(\ell+1)}}
\bar\rho^\ell_{ij}(p).
\nn
\end{equation}
We find
\begin{equation}
F_{\ell}^{ij;km}(p)=\left(\pi^{ik}\, \pi^{jm}+\pi^{im}\, \pi^{jk}-
{2\over 5}\pi^{ij}\, \pi^{km}\right){2R^2 \, \chi_\ell(p)}.
\nn
\end{equation}
This correlator is analytic in the infra-red.
Thus, the presence of the shell seems to
completely break the conformal symmetry.

\boldmath
\section{The spectrum of massless localized states}\label{jspec}
\unboldmath

In this section we will investigate the spectrum of massless states
localized
on the brane distribution at $r=R$. The correct but tedious procedure is to
study in detail the fluctuations of the background solution including the
$\delta$-function sources. We will however take here a short-cut due to the
fact that the background configuration is BPS, and the massless fluctuations
are constrained by lack-of-force condition and supersymmetry. They
correspond therefore to deformations of the brane distribution
plus supersymmetric patterns.

It is therefore enough to consider the background ``BPS condition",
Eq. (\ref{boxdelta}):
\begin{equation}
\mysquare_{4} h=-2{N \over R^3}\, \delta(r-R),
\nn
\end{equation}
where we have assumed that $h$ may depend in general on all four transverse
coordinates $\{y^a, a=6, \ldots, 9\}\equiv \{r,\Omega \}$
($\mysquare_{4}$ is the corresponding ``flat'' Laplacian
introduced previously). Our aim is to go beyond the spherical solution of
$N$ five-branes, Eq. (\ref{hsing}).
Perturbations of this solution involve deformations of the shell as well as
of
the charge density, keeping the total charge fixed. Thus, the equation for
the small fluctuations is
\begin{equation}
\mysquare_{4} (h+\delta h) =-4\pi^2
\left(\varrho_0 + \delta\varrho(\Omega)\right)
\delta(r-R + \delta f(\Omega)),
\label{fluct}
\end{equation}
where $\varrho_0 = {N\over 2\pi^2 R^3}$
is the unperturbed constant density of five-branes over the three-sphere,
and $\delta\varrho(\Omega)$ and $\delta f(\Omega)$ small perturbations
of the density and shape of the distribution.
Expanding Eq. (\ref{fluct}) to first order we
obtain\footnote{This expansion is formal. Strictly speaking it is valid
away from $r=R$.}:
\begin{equation}
\mysquare_{4} \delta h=
-4\pi^2  \delta\varrho(\Omega)\, \delta(r-R)
-4\pi^2  \varrho_0 \, \delta f(\Omega)\, \delta'(r-R).
\label{fluctexp}
\end{equation}

In the rest of this section, we will try to get a flavor of the massless
spectrum as it comes out of Eq. (\ref{fluctexp}).
Let us expand the various functions in $SU(2)$ spherical harmonics:
\begin{eqnarray}
\delta h(r,\Omega)&=&\sum_{\ell,n,n'}
h_{\ell}^{n,n'}(r)\,
D^\ell_{n,n'}(\Omega),\nn \\
\delta\varrho(\Omega)&=&-{1\over 4\pi^2} \sum_{\ell,n,n'}
A_{\ell}^{n,n'}\,
D^\ell_{n,n'}(\Omega),\nn \\
\delta f(\Omega)&=&-{1\over 4\pi^2\, \varrho_0} \sum_{\ell, n,n'}
B_{\ell}^{n,n'}\,
D^\ell_{n,n'}(\Omega),\nn
\end{eqnarray}
and introduce them into Eq. (\ref{fluctexp});
we obtain the decoupled equations\footnote{As usual,
$-\ell\leq n,n'\leq \ell$. We
will suppress the $n,n'$ quantum numbers from now on.}:
\begin{equation}
h_\ell''+{3\over r}\, h_\ell'-{\ell(\ell+2)\over r^2}\, h_\ell=
A_\ell\, \delta(r-R)+B_\ell \, \delta'(r-R).
\label{fluctexp2}
\end{equation}
The condition that the total charge is $N$ translates into
\begin{equation}
RA_0-3B_0=0.
\label{sumrule}
\end{equation}
The regular solutions to Eqs. (\ref{fluctexp2}) at $r\not=R$ are given by
\begin{equation}
h_\ell(r)=\left\{ \begin{array}{ll}
\displaystyle a_\ell\, r^\ell, &0\leq r<R,\\
\\
\displaystyle b_\ell \, r^{-\ell-2},&R<r.
\end{array}\right.
\label{22}
\end{equation}

We can go further by matching the $\delta$-functions. This leads to the
following set of relations:
\begin{equation}
R^\ell\, a_\ell =
{R\, A_\ell+(\ell - 1)B_\ell\over 2(\ell+1)}\ , \ \
R^{-\ell-2}\, b_\ell =
{R\, A_\ell-(\ell + 3) B_\ell\over 2(\ell+1)}.
\nn
\end{equation}
In the case $\ell=0$, e.g.,  taking into account the charge neutrality
condition (\ref{sumrule}) we find:
\begin{equation}
h_0(r)=\left\{ \begin{array}{ll}
\displaystyle B_0, &0<r<R,\\
\\
\displaystyle 0,&R<r,
\end{array}\right.
\nn
\end{equation}
which is normalizable. The other solutions (\ref{22}) can be analyzed
similarly. Furthermore, a fluctuation analysis of the effective action again
indicates that such localized modes will have finite six-dimensional
couplings
if the norm of their wave functions, defined as
\begin{equation}
\langle \delta h |\delta h\rangle=\int d^4 y\,
 h(r)\, (\delta h(y))^2,
\nn
\end{equation}
is finite ($d^4y$ is the ``flat'' transverse measure).
It is obvious from the behavior of the wave functions above that all are
normalizable. This completes the determination of the spectrum of the bulk
and  massless moduli.

A last comment is in order here. The continuous and smooth distributions
of branes we
are considering are good approximations when $N\to\infty$. In particular,
the spherical distribution at hand, can be generated by putting
$N$ single
branes uniformly on $S^3$. The mean distance between nearest neighbors  then
scales like  $R\,\ell_{\rm s}/ N^{1/3}$.
This implies that the corresponding cut-off in
the angular momenta is $\ell< \ell_0 \sim N^{1/3}$.
If we attempt a counting of the massless modes we obtain:
\begin{equation}
 {\rm \# \  of \ moduli}=
4\left(1 + 2 \sum_{\ell=1}^{\ell_0} (2\ell+1)^2\right) =
{8\over 3}\, \ell_0\, (2\ell_0+1)\, (4\ell_0+1)-4
\sim {\cal O}(N),
\nn
\end{equation}
in qualitative agreement with the exact answer $4N$.

\boldmath
\section{Conclusions and further problems}
\unboldmath

We have investigated five-brane distributions, that have the property that the
strong-coupling region is absent, and they have high symmetry so that
detailed
calculations become possible. A characteristic feature
of the distributions we have
studied is the appearance of a ``discontinuous'' geometry, and therefore
the need for including
sources. However, as we have explicitly shown, those are controllable
backgrounds, and the study of small fluctuations around them is
well defined.

We have considered distributions that correspond to exact CFT's
albeit of a new kind. They correspond to sowing together (in
space--time) known CFT's. The simplest example is a spherical shell of
$N$
NS5-branes distributed uniformly on an $S^3$ in transverse space.
The number $N$ is assumed to be large enough
so that the geometry is weakly curved, and $\a'$ corrections to
supergravity
negligible. The brane distribution can be approximated by
a continuous one, and therefore enjoy high symmetry ($SO(4)$).

In the interior of the shell the geometry and other background fields are
flat.
There are five-brane $\delta$-function sources at the position of the shell.
We have shown that the background fields are determined uniquely
from the supergravity equations.
The radius of the shell $R$ can be chosen large, $R\gg\sqrt{N}$, so that the
string coupling is weak outside the shell. Inside the shell the string
coupling is frozen. Thus, there is no strong-coupling region in such a
background.

We have also described a richer spectrum of such backgrounds using also
negative-tension branes. In the case of the D5-branes these are no-other
than
the orientifold five-planes. For NS5-branes, their
negative-tension cousins are
``bare" $Z_2$ orbifold five-planes. A usual orbifold five-plane appearing as a
twisted sector in closed string orbifold vacua is a bound state of an
NS5-brane
and a bare orbifold plane that cancels the tension and charge of the
NS5-brane
much alike the situation in orientifold vacua.

A special configuration (in type-IIB context) in this sense is one where in
the interior section is a D5-brane while asymptotically it is an NS5-brane.
The two configurations match on a shell of NS5-branes and O5-planes. No
strong-coupling region exists also here.

We have also studied the spectrum of fluctuating fields. These correspond
via holography to operators of the boundary LST. The effective field theory of
such fluctuations is expected to be a seven-dimensional $SO(4)$ gauged
supergravity.
It is obtained by compactifying the ten-dimensional type-IIA/B supergravity
(in the string frame) on $S^3$ with the appropriate parallelizing
flux of the
antisymmetric tensor. The near-horizon region of
an NS5-brane corresponds to a flat seven-dimensional
space and a linear dilaton in
one direction. This is  the holographic direction.

We have explicitly solved for the fluctuations of some of the fields of the
bulk theory. These include the six-dimensional graviton (corresponding to
the boundary stress tensor) and its associated K--K tower.
It turns out that the six-dimensional graviton satisfies an
equation without sources. We have found
the normalizable modes and shown that its spectrum has a mass gap
$\sim {{\ell+1}\over \sqrt{N}}M_{\rm s}$.
This in accordance with an earlier CFT
computation \cite{AFK,KK}.

The other set of fluctuations we have considered are the moduli modes which
are
massless. These satisfy a Laplace equation with sources. The sources are
crucial for the existence of normalizable moduli modes as we have shown.

We have further  studied the non-normalizable modes of the six-dimensional
graviton in order to apply the holographic principle. The symmetries of the
background  are $SO(4)\times{ISO(6)}$. The $SO(4)$ corresponds
to the R-symmetry of
the boundary theory, while the rest is the usual Euclidean group in six flat
dimensions. This is unlike AdS-like spaces, where also conformal
transformations
are boundary symmetries. The reason is that here the boundary theory is
massive.

Using the bulk supergravity action we have computed the boundary two-point
function of two stress tensors in the stringy (LST) regime. We can remind its
features:

\noindent
(\romannumeral1) its long distance behavior is massive with associated mass
${M_{\rm s}\over \sqrt{N}}$;

\noindent
(\romannumeral2) in the formal $N\to\infty$ limit it becomes power-like with a
$|x|^{-7}$ behavior;

\noindent
(\romannumeral3) the stress tensor has canonical mass dimension 7 due to a
non-trivial IR wave-function renormalization of its source;

\noindent
(\romannumeral4) it is independent of
the presence of the shell and, as we argue, this is no longer true for
higher correlators.

One the other hand, in the effective-field-theory regime, a renormalized
stress tensor
exists with correlators that depend on the presence of the shell, but it
analytic at low
momenta implying that conformal invariance is completely broken.

There are several problems that require further study in relation with the
approach described in this paper.

The full structure and spectrum of fluctuations around these   supergravity
backgrounds should be worked out. As for the spectrum, it can be obtained
directly by an $S^3$ compactification of the ten dimensional supergravity.
Finding however the interactions of the massless fields might require a
direct
approach of gauging the $SO(4)$ group in a seven-dimensional reduction around
flat space. This would be essential for performing further concrete
calculations of boundary correlation functions.

A further effort is needed in order to eventually interpret the supergravity
results and elucidate the physics of the boundary LST. Especially in the
type-IIA case, this is hampered by the lack of a useful low-energy effective
description that can be used to interpret the supergravity results. The hope
is
that the holographic amplitudes might suggest a useful and transparent such
description.

String corrections to the supergravity results will be useful to calculate.
As
mentioned earlier this will entail solving a new kind of CFT, namely one
that
has fixed walls in space--time (and which translates into fluctuation
boundaries
on the world-sheet). Preliminary investigation indicates that such a CFT
involves D-branes with fluctuating boundaries. It will be very
interesting, and potentially useful to understand such CFT's.

\acknowledgments

The ideas of the present work were triggered during the IHP session on
string theory (Paris 2000--2001).
We would like to thank O. Aharony, C. Angelantonj, C. Bachas, M.
Berkooz, M. Bianchi and A. Petkou  for discussions.
We all thank each-other's institute for hospitality during various stages
of the project.
We would also like to thank the referee for
constructive criticism.
This work was partially
supported by European Union under the RTN
contracts HPRN-CT-2000-00148, HPRN--CT--2000--00122 and HPRN--CT--2000--00131.
The work of E. K. was also partially supported by Marie Curie
contract MCFI-2001-0214.

\appendix

\section{Bulk-to-bulk propagator}
We will compute in this appendix the bulk-to-bulk propagator in the
background corresponding to the spherical shell of NS5-branes given in
(\ref{hsing}), albeit in Euclidean setting.
There are four distinct cases,
$B_\ell^{\pm \pm} (x,r;x',r')$, depending on whether the variables $r$
and $r'$ lie in $[0,R]$ (minus sign) or $[R,\infty[$ (plus sign).

The Fourier transform,
\begin{equation}
B_\ell^{\pm \pm}(x,r;x',r')={1\over {\alpha'}^3}\int{d^6p\over (2\pi)^6}
\e^{i p \cdot (x-x') }\, B_\ell^{\pm \pm}(p;r,r'),
\nn
\end{equation}
satisfies (for each partial wave $\ell$) the following equation:
\begin{equation}
\begin{split}
\left[{1\over r^3}\partial_rr^3\partial_r-{\ell(\ell+2)\over r^2}-
{Np^2\over r^2} \right]B^{++}_\ell(p;r,r')
=&{1\over r^3}\delta(r-r')\ , \ \
r,r'\in[R,\infty[,\\
\left[{1\over r^3}\partial_rr^3\partial_r-{\ell(\ell+2)\over r^2}-
{Np^2\over r^2} \right]B^{+-}_\ell(p;r,r')
=&0\ , \ \
r\in[R,\infty[\ , \ \  r'\in[0,R],\\
\left[{1\over r^3}\partial_rr^3\partial_r-{\ell(\ell+2)\over r^2}-
{Np^2\over R^2} \right]B^{-+}_\ell(p;r,r')
=&0\ , \ \
r'\in[R,\infty[\ , \ \  r\in[0,R],\\
\left[{1\over r^3}\partial_rr^3\partial_r-{\ell(\ell+2)\over r^2}-
{Np^2\over R^2} \right]B^{--}_\ell(p;r,r')
=&{1\over r^3}\delta(r-r')\ , \ \
r,r'\in[0,R].
\end{split}
\nn
\end{equation}

We impose the following conditions: regularity in variables $r$
and $r'$ both at
zero and infinity, as well as continuity of the propagator and its first
derivatives across the shell. The answer is unique:
\begin{equation}
B^{++}_\ell(p;r,r')=\left\{ \begin{array}{lc}
\displaystyle{{R^{2a}\over 2a}\,
{bRI'_{\ell+1}(bR)-aI_{\ell+1}(bR)\over
bRI'_{\ell+1}(bR)+aI_{\ell+1}(bR)}\,
(rr')^{-1-a}}, &r>r',\\
\\
\displaystyle{{R^{2a}\over 2a}\,
{bRI'_{\ell+1}(bR)-aI_{\ell+1}(bR)\over
bRI'_{\ell+1}(bR)+aI_{\ell+1}(bR)}\,
(rr')^{-1-a}+{r^{-a}{r'}^{a}-r^{a}{r'}^{-a}\over
2arr'}} ,&r<r',
\end{array}\right.
\nn
\end{equation}
\begin{equation}
B^{+-}_\ell(p;r,r')=-{R^a I_{\ell+1}(br')\over
bRI'_{\ell+1}(bR)+aI_{\ell+1}(bR)}\,
r^{-1-a},
\nn
\end{equation}
\begin{equation}
B^{-+}_\ell(p;r,r')=-{R^a I_{\ell+1}(br)\over
bRI'_{\ell+1}(bR)+aI_{\ell+1}(bR)}\,
{r'}^{-1-a},
\nn
\end{equation}
\begin{equation}
B^{--}_\ell(p;r,r')=\left\{ \begin{array}{lc}
\displaystyle{\left({bRK'_{\ell+1}(bR)+aK_{\ell+1}(bR)\over
bRI'_{\ell+1}(bR)+aI_{\ell+1}(bR)}I_{\ell+1}(br)-K_{\ell+1}(br)\right)}
I_{\ell+1}(br'), &r>r',\\
\\
\displaystyle{\left({bRK'_{\ell+1}(bR)+aK_{\ell+1}(bR)\over
bRI'_{\ell+1}(bR)+aI_{\ell+1}(bR)}I_{\ell+1}(br')-K_{\ell+1}(br')\right)}
I_{\ell+1}(br),&r<r',
\end{array}\right.
\nn
\end{equation}
where
\begin{equation}
 a=\sqrt{(\ell+1)^2+Np^2}\ , \ \  b={\sqrt{p^2 N}\over R}.
\nn
\end{equation}

Note that the bulk-to-bulk propagator is free of poles inside momentum
space. Moreover, its momentum-dependent coefficients go fast to zero at large
momenta. In Minkowski space, this propagator is still valid below the mass
gap.
Above the mass gap the eigenfunctions of Section 7 should be used.

\end{document}